\documentclass[preprint2]{aastex}
\usepackage{rotating}
\usepackage{graphicx}
\usepackage{grffile}

\begin{document}
\setcounter{secnumdepth}{2}     
\title{The TRENDS High-Contrast Imaging Survey. I. \\ Three Benchmark M-Dwarfs Orbiting Solar-type Stars}
\author{Justin R. Crepp\altaffilmark{1,2}, John Asher Johnson\altaffilmark{1}, Andrew W. Howard\altaffilmark{3,4}, Geoff W. Marcy\altaffilmark{3}, Debra A. Fischer\altaffilmark{5}, Lynne A. Hillenbrand\altaffilmark{1}, Scott M. Yantek\altaffilmark{1}, Colleen R. Delaney\altaffilmark{1}, Jason T. Wright\altaffilmark{6,7}, Howard T. Isaacson\altaffilmark{3}, Benjamin T. Montet\altaffilmark{1}}
\altaffiltext{1}{Department of Astronomy, California Institute of Technology, 1200 E. California Blvd., Pasadena, CA 91125, USA} 
\altaffiltext{2}{Department of Physics, University of Notre Dame, 225 Nieuwland Science Hall, Notre Dame, IN 46556, USA} 
\altaffiltext{3}{Department of Astronomy, University of California, Berkeley, CA 94720, USA} 
\altaffiltext{4}{Institute for Astronomy, 2680 Woodlawn Drive, Honolulu, HI 96822-1839, USA}
\altaffiltext{5}{Department of Physics, Yale University, New Haven, CT 06511, USA} 
\altaffiltext{6}{Department of Astronomy \& Astrophysics, The Pennsylvania State University, University Park, PA 16802, USA} 
\altaffiltext{7}{Center for Exoplanets and Habitable Worlds, The Pennsylvania State University, University Park, PA 16802, USA}
\email{jcrepp@nd.edu} 

\begin{abstract}  % Submit to ApJ
We present initial results from a new high-contrast imaging program dedicated to stars that exhibit long-term Doppler radial velocity accelerations (or ``trends"). The goal of the TRENDS ({\bf T}a{\bf R}getting b{\bf EN}chmark-objects with {\bf D}oppler {\bf S}pectroscopy and) imaging survey is to directly detect and study the companions responsible for accelerating their host star. In this first paper of the series, we report the discovery of low-mass stellar companions orbiting HD~53665, HD~68017, and HD~71881 using NIRC2 adaptive optics (AO) observations at Keck. Follow-up imaging demonstrates association through common proper-motion. These co-moving companions have red colors with estimated spectral-types of K7--M0, M5, and M3--M4 respectively. We determine a firm lower-limit to their mass from Doppler and astrometric measurements. In the near future, it will be possible to construct three-dimensional orbits and calculate the dynamical mass of HD~68017~B and possibly HD~71881~B. We already detect astrometric orbital motion of HD~68017~B, which has a projected separation of $13.0$ AU. Each companion is amenable to AO-assisted direct spectroscopy. Further, each companion orbits a solar-type star, making it possible to infer metallicity and age from the primary. Such benchmark objects are essential for testing theoretical models of cool dwarf atmospheres.
\end{abstract}

\section{INTRODUCTION}\label{sec:intro}
% Is the radial velocity technique sensitive to distant companions? Yes. 
It is commonly thought that the radial velocity (RV) method for detecting companions to nearby stars only provides information about low-mass bodies with short orbital periods. However, Doppler measurements are quite sensitive to distant objects, because the RV semi-amplitude, $K$, decreases slowly with orbital period, $P$ (or semimajor axis, $a$), according to: 
\begin{eqnarray}
K &\propto& m \: \sin(i) \: P^{-1/3} \\
    &\propto& m \: \sin(i) \: a^{-1/2}
\end{eqnarray}
where $m$ is the companion mass and $i$ is the orbit inclination. With $\sim$1-10 m/s precision, M-dwarfs, brown dwarfs, and super-Jupiters are detectable out to tens of AU, it just takes a long time to complete a full orbit (e.g., \cite{howard_10}). Despite having knowledge of only a fraction of an orbital cycle, RV accelerations (trends) are tremendously useful: they show us conclusively that something initially hidden from view is tugging on the visible star. 

% Why are high efficiency observations important? 
Informed target selection is an order of magnitude effect for substellar companions. For example, wide-separation ($a \gtrsim 10$ AU) brown dwarfs are rare, having been found to orbit only $\approx$3\% of solar-type stars \citep{metchev_hillenbrand_09}. And, while companions in the planetary-mass regime are expected to be more common, an upper-limit to the frequency of super-Jovian ($m\gtrsim3M_J$) bodies over a similar semi-major axis range is set at $\approx$20\% for solar-type stars \citep{nielsen_10,vigan_12}. By observing a sample of intrinsically companion-rich sources, those with clear Doppler accelerations, it is possible to by-pass the inefficiencies common to high-contrast programs that nominally select stars based solely on age and proximity to the Sun \citep{masciadri_05,biller_07,lafreniere_07,leconte_10,ehrenreich_10}. 

% What is the science motivation? 
In addition to high observing efficiency, there are significant scientific benefits to combining Doppler measurements with high-contrast imaging. RV and direct astrometry observations\footnote{By direct astrometry we are referring to following the companion along its orbit with direct imaging, measuring the sky-projected separation and position angle relative to the star as a function of time. Indirect astrometry involves measuring the position of the star relative to other (distant) nearby stars or fiducial reference points.} may be used in concert to calculate the companion orbit (all 6 elements) and dynamical mass \citep{boden_06}.\footnote{See \citealt{rodigas_11} for a discussion regarding mass and orbit constraints for the case of Doppler accelerations combined with direct imaging non-detections.} For example, \citealt{crepp_12a} have measured the mass of the benchmark brown dwarf HR~7672~B with a fractional uncertainty of $4\%$ by monitoring its motion over $\sim$33$\%$ of an orbit cycle. Masses determined independent of photometry and spectroscopy inform theoretical atmospheric models by helping to break degeneracies between the various input parameters, such as mass, radius, age, effective temperature, and chemical composition \citep{barman_11,janson_11}. Dynamical masses are likewise important for calibrating thermal evolutionary models, providing a measure of substellar objects luminosity as they fade with time \citep{stevenson_91,burrows_97}. Further, if the companion orbits a solar-type star, the metallicity and age may be inferred from the primary \citep{liu_07,dupuy_09,johnson_09,crepp_12a,biller_10_pztel}. 

% What is our program about? 
With this motivation in mind, we have established an interdisciplinary program that uses a combination of Doppler observations and high-contrast imaging. Specifically, we use years of precise RV measurements to identify promising targets for follow-up high-contrast observations. The goals of our program are to:
\begin{enumerate}
\item{detect companions with mass in the $m\approx5M_J-500M_J$ range;}
\item{acquire spectro-photometric measurements across a wide bandpass (YJHKLM);}
\item{perform follow-up direct astrometric measurements to break the $\sin \: i$ inclination degeneracy resulting from Doppler measurements and calculate dynamical masses;}
\item{test theoretical atmospheric and evolutionary models in a regime where they currently break-down (low temperatures).}
\end{enumerate}

% Where does the data come from? What about Lick and Palomar? 
Based primarily at Keck Observatory, our RV measurements are obtained using the HIgh Resolution Echelle Spectrometer (HIRES; \citealt{vogt_94}) at Keck I. These measurements are often augmented by previous and concurrent observations at Lick Observatory. High-contrast imaging observations are obtained using NIRC2 (PI: Keith Matthews) and the Keck II AO system \citep{wizinowich_00}. We are also expanding this program to include AO observations with the Project 1640 spectral-imager at Palomar \citep{hinkley_11_PASP} and LMIRCam at the Large Binocular Telescope \citep{skrutskie_10,skemer_12}. Our strategy is to maximize on-sky sensitivity by employing all of the powerful techniques recently developed for high-contrast imaging applications, including coronagraphy \citep{guyon_06,crepp_10} and aggressive point-spread function subtraction to remove residual scattered starlight from images \citep{marois_06,lafreniere_07,crepp_11}. 

% What is this paper about? 
We have found that selecting targets based on the presence of an RV trend is an effective approach, allowing one to take a ``short-cut" for finding stars likely to host a low-mass companion amenable to direct imaging detection. In this paper, we report the the discovery of three M-dwarf companions orbiting solar-type stars. Each companion has red colors and a low luminosity, otherwise they would be noticed as double-lined spectroscopic binaries in RV data. The companions are amenable to AO-assisted spectroscopy \citep{bowler_10_hr8799b,pueyo_12} and represent the first discoveries of our program.

%High-mass ratio binaries are less common than equal-mass binaries (Wisniewski et al. 2012) ...

\section{OBSERVATIONS}

\subsection{High-Resolution Stellar Spectroscopy}
\subsubsection{Doppler Measurements}
Precise RV data were obtained with the HIgh-Resolution Echelle Spectrometer (HIRES; \citealt{vogt_94}) at Keck. We use the iodine cell referencing method to calibrate instrument drift and measure Doppler shifts \citep{marcy_butler_92,butler_96}. Observations for HD~53665, HD~68017 and HD~71881 began on Jan. 25, 1998, Jan. 13, 1997, and Dec. 24, 1997 respectively. In each case, a long-term acceleration indicated that the star is orbited by a distant body (Fig. 1). Accelerations are approximately linear for HD~53665 and HD~71881, whereas HD~68017 shows significant orbit curvature (change in the acceleration). In $\S$\ref{sec:dyn}, we use Doppler measurements in combination with imaging observations to constrain the mass of each companion. 
%Doppler measurements and uncertainties are listed in Appendix A (insufficient room for a Letter). 
% We calculate lower-limits for the dynamical mass of their companions based on the trend ($\S$\ref{sec:sptp_mass}).

\begin{figure}[!t]
\begin{center}
\includegraphics[height=6in]{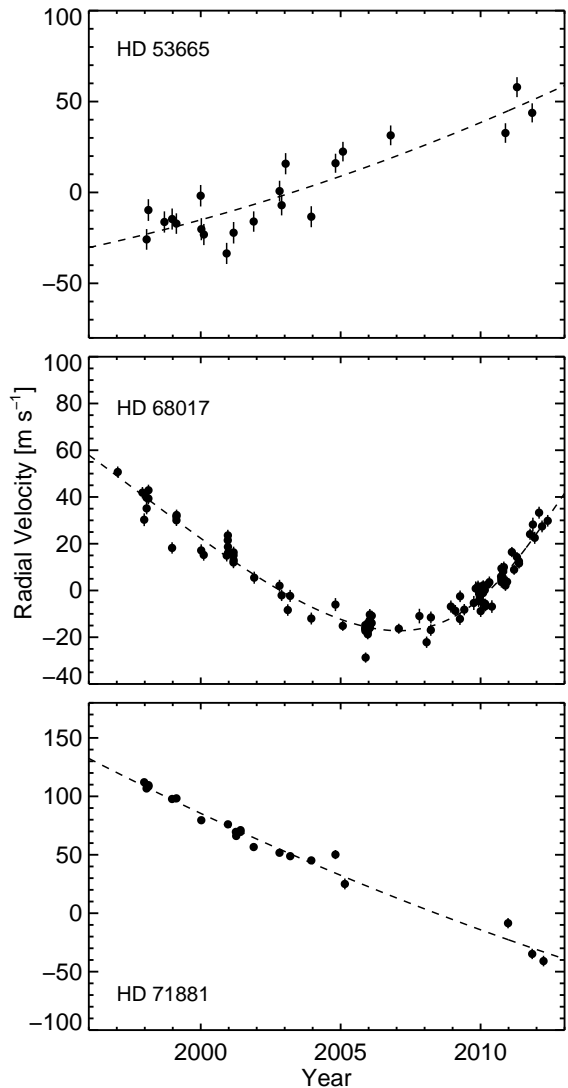} 
\caption{Precise Doppler radial velocity measurements. Our time baseline exceeds a decade for each star. HD~68017 shows significant curvature. It will be possible to calculate the dynamical mass of HD~68017~B with only several more astrometric measurements.} 
\end{center}\label{fig:rvs}
\end{figure} 

\subsubsection{Stellar Properties}
Stellar (template) spectra, taken with the iodine gas cell removed from the optical path, were analyzed using the LTE spectral synthesis code {\it Spectroscopy Made Easy} (SME) described in \citealt{valenti_fischer_05}. SME provides an estimate of the stellar effective temperature ($T_{\rm eff}$), surface gravity ($\log g$), metallicity ($\mbox{[Fe/H]}$), and projected rotational velocity ($v \sin i$). Table ~\ref{tab:starprops} lists the spectral-type and physical properties of each star derived from spectral fitting along with comparison to theoretical isochrones. 

\begin{figure*}[!t]
\begin{center}
\includegraphics[height=2.1in]{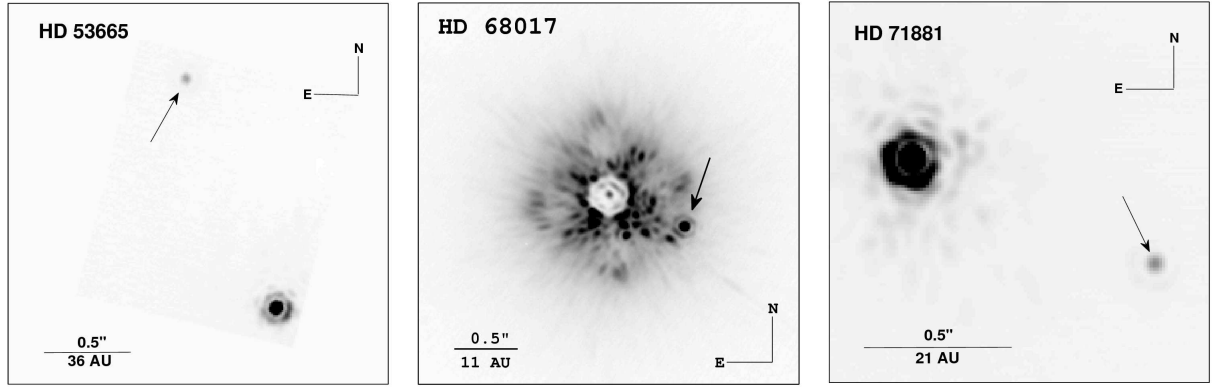} 
\caption{Discovery images of HD~53665~B, HD~68017~B, and HD~71881~B taken in the K' filter ($\lambda_c=2.12 \; \mu$m) with NIRC2 on Feb. 22, 2011. Companions are indicated by an arrow.} 
\end{center}\label{fig:images}
\end{figure*} 

\subsection{Adaptive Optics Imaging}
HD~53665, HD~68017, HD~71881 were each discovered on the same night, February 22, 2011 UT, using NIRC2 at Keck. We used the narrow camera setting ($9.963\pm0.006$ mas pix$^{-1}$ plate-scale \citep{ghez_08}) to provide fine spatial sampling of the system point-spread function. Images were initially obtained in the $K'$ ($\lambda_c=2.12 \mu$m) filter for search-mode operation. Companions to each star were noticed in raw frames. We executed a three-point dither pattern to facilitate removal of sky background noise. The companion orbiting HD~68017 is fainter and closer to its host star compared to the other targets, so we also obtained images placing the star behind the (partially transmissive) 300 mas diameter coronagraphic mask. 

Images were processed by flat-fielding, correcting for hot pixels with interpolation, subtracting the sky background, and rotating the frames to standard north-east orientation. Fig.~2 shows fully processed images of each companion. We acquired follow-up observations on January 7, 2012 UT in different filters to obtain colors and assess whether the candidates were associated with their host star. The angular separation and position angle of each companion are listed in Table~1. Brightness ratios are listed in Table~2. 

\begin{table*}[!t]
\centerline{
\begin{tabular}{lccccc}
\hline
\hline
 Target Name      &   Date (UT)    &   JD-2,450,000         &      $\rho$ (mas)      &    P. A. (degrees)    & Proj. Sep. (AU)    \\
\hline
\hline        
HD~53665~B         &    Feb. 22, 2011       &    5614.79          &      $1420.7\pm1.0$     &    $21.3^{\circ}\pm0.1^{\circ}$   &  $102.9^{+5.1}_{-6.0}$ \\
                               &    Jan. 7, 2012          &    5933.93          &      $1413.4\pm0.9$     &    $20.9^{\circ}\pm0.1^{\circ}$    &  $102.4^{+5.0}_{-6.1}$ \\
HD~68017~B         &    Feb. 22, 2011        &    5614.81         &      $594.5\pm0.5$       &    $248.2^{\circ}\pm0.1^{\circ}$  &   $13.0^{+0.1}_{-0.2}$ \\
                               &    Jan. 7, 2012          &    5933.96          &      $574.6\pm0.5$       &    $240.3^{\circ}\pm0.1^{\circ}$  &   $12.5^{+0.1}_{-0.2}$ \\
HD~71881~B         &    Feb. 22, 2011        &   5614.81          &      $851.7\pm1.1$       &    $247.8^{\circ}\pm0.1^{\circ}$   &   $35.2^{+0.9}_{-1.1}$ \\
                               &    Jan. 7, 2012          &    5933.97          &      $859.3\pm0.6$       &    $246.7^{\circ}\pm0.1^{\circ}$   &  $ 35.5^{+0.9}_{-1.1}$ \\
\hline
\hline
\end{tabular}}
\caption{Summary of astrometric measurements.}
\label{tab:astrometry}
\end{table*}

\section{ASTROMETRY}
Our astrometric observations consist of two epochs separated by 0.9 years for each source (Table~2). All three stars have large proper-motions (see Table 1), allowing us to easily determine whether the companions share the same space motion over this time frame. We measured an accurate separation and position angle of each companion following the technique described in \citealt{crepp_12a}. We first fit two-dimensional Gaussian functions to the stellar and companion point-spread functions to locate their centroids in each frame. The primary star was not saturated in any of our dithered images. We then correct for distortion in the NIRC2 focal plane using publicly available solutions provided by Keck Observatory's astrometry support page \footnote{http://www2.keck.hawaii.edu/inst/nirc2/forReDoc/post$\_$observing/dewarp/}. The results are averaged and uncertainty in the separation and position angle is taken as the standard deviation, taking into account uncertainty in the plate scale and orientation of the array by propagating these errors to the final calculated position. 

\begin{figure*}[!t]
\begin{center}
\includegraphics[height=2.0in]{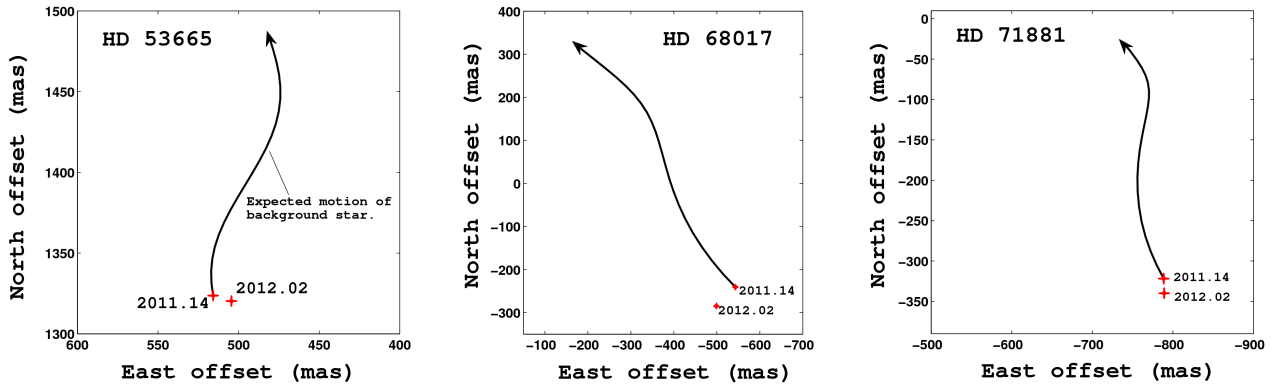} 
\caption{Astrometric measurements (red-crosses) demonstrating that each companion is comoving with its host star. Solid-curves show the path that a distant background object would execute over the same time frame accounting for proper-motion and parallactic motion from Feb. 22, 2011 through Jan. 7, 2012.} 
\end{center}\label{fig:astrometry}
\end{figure*} 

Fig.~3 shows multi-epoch astrometry measurements plotted against the expected motion of a distant background object. We find that all three companions, HD~53665~B, HD~68017~B, and HD~71881~B, are each clearly associated with their respective host stars, implying that they are gravitationally bound. The positions of HD~53665~B and HD~71881~B have changed by $\approx$1-2 pixels over the 0.9 year time-frame. Meanwhile, an unrelated background source placed at infinite distance would have moved relative to the host star by 163.7 mas (16.4 pix) and 296.5 mas  (29.8 pix) respectively. HD~68017~B has the smallest projected separation of the three ($13.0^{+0.1}_{-0.2}$ AU, Feb. 2011) and appears to exhibit significant orbital motion in a clockwise direction. 

\begin{table*}[!t]
\centerline{
\begin{tabular}{lccc}
\hline
\hline
                                         &   HD 53665 A &    HD 68017 A   &    HD 71881 A    \\
\hline
\hline
R.A. (J2000)                    &  07 05 52.8    &    08 11 38.6      &  08 31 55.0            \\
Decl. (J2000)                   &  -01 01 13.7   &  +32 27 25.7     &  +50 37 00.1          \\
$B$ (mag)                         &   $7.76\pm0.02$        &   7.50            &    8.06                        \\
$V$ (mag)                        &    $7.26\pm0.01$        &   6.81            &    7.43                        \\
$J_{\rm 2MASS}$ (mag)     &   $6.294\pm0.024$     &  5.48             &   $6.284\pm0.020$   \\
$H_{\rm 2MASS}$ (mag)    &   $6.066\pm0.027$    &  5.15             &   $6.027\pm0.027$    \\
$K_{\rm 2MASS}$ (mag)    &   $5.983\pm0.018$    &  5.09              &   $5.959\pm0.029$    \\
d (pc)                                 &   $72.2_{-4.0}^{+3.9}$  &  $21.8_{-0.3}^{+0.2}$   &   $41.3_{-1.2}^{+1.1}$    \\
p.m. (mas/yr)                    &   9.4, -15.1                     & -462.6, -644.2         &  -81.4, -338.6     \\
Spec. Type                       &    F8V                             &     G4V                        &     G1V                    \\
$M_* (M_{\odot})$             &      $1.51^{+0.20}_{-0.27}$  &    $0.85^{+0.04}_{-0.03}$       &   $1.04^{+0.06}_{-0.05}$     \\
$T_{\rm eff}$ (K)                    &   $ 6225  \pm 44$          &    $5552\pm44$       &  $5821\pm44$        \\
log g (cm/$s^2$)            &   $4.05 \pm0.06$           &    $4.65\pm0.06$      &   $4.29\pm0.06$        \\
$[\mbox{Fe/H}]$             &   $0.17\pm0.04$            &   $-0.44 \pm 0.03$    &   $-0.05\pm0.03$        \\
$v \sin i$ (km/s)              &   $8.6 \pm 0.5$              &     $0.8\pm0.5$         &   $2.1\pm0.5$        \\
$\log R'_{HK}$                & $-5.022\pm0.009$        & $-4.928\pm0.013$   &   $-5.043\pm0.002$ \\
%Age (Gyr) -- isochrones from SPOCS website  & $2.1^{+0.8}_{-0.4}$     &  $11.0^{+2.7}_{-5.5}$     &  $6.4^{+1.6}_{-3.4}$   \\ 
Age (Gyr)                      &     $2.1^{+0.8}_{-0.4}$     &    $4.6^{+0.9}_{-1.0}$    &   $4.3^{+1.0}_{-1.2}$    \\
\hline
\\
\hline
\hline
                             &   HD 53665 B    &   HD 68017 B    &   HD 71881 B    \\
\hline
\hline        
$\Delta J_{\rm MKO}$        &    $3.85\pm0.15$                  &   ---                             &   ---       \\
$\Delta H_{\rm MKO}$       &    $3.26\pm0.07$                  &  $> 4.16$       &   $4.26\pm0.04$       \\
$\Delta K_{\rm MKO}\approx \Delta K'_{\rm MKO}$       &     $3.14\pm0.10$                 &   $4.92\pm0.10$          &    $4.12\pm0.08$      \\
$J_{\rm MKO}$                       &   $10.12\pm0.16$                &   ---                                &   ---             \\
$H_{\rm MKO}$                      &   $9.31\pm0.08$                  &   $>9.29$                     &   $10.26\pm0.07$          \\
$K_{\rm MKO}$                      &    $9.11\pm0.11$    &  $10.00\pm0.11$       &    $10.07\pm0.09$    \\ 
% have added 0.01 mags to account for different K-filter
$M_{J_{\rm MKO}}$                  &   $5.82\pm0.20$               &     ---                            &    ---      \\
$M_{H_{\rm MKO}}$                 &    $5.01\pm0.14$              &   $>7.60$                    &  $7.18\pm0.09$        \\
$M_{{K'}_{\rm MKO}}$              &   $4.81\pm0.16$              &   $8.31\pm0.11$          &     $6.99\pm0.11$        \\
Spec. Type                             &         K7--M0                           &     M5                      &     M3--M4     \\
$m_{dyn}$ ($M_{\odot}$)       &    $>0.63$         &    $>0.08$                 &      $>0.17$    \\
% $m_{phot}$ ($M_{\odot}$)  &   $0.61\pm0.03$   &  $0.15\pm0.01$        &   $0.28\pm0.02$   \\ Girardi et al. 2002.
$m_{\rm model}$ ($M_{\odot}$)       &   $0.65\pm0.03$   &  $0.16\pm0.02$        &   $0.31\pm0.03$   \\  % Dotter et al. 2008 theoretical models
$m_{\rm empirical}$ ($M_{\odot}$)  &   $0.70\pm0.03$   & $0.15\pm0.01$     &     $0.29\pm0.02$    \\ % Delfosse et al. 2000 M_K empirical fits
\hline
\end{tabular}}
\caption{(Top) Coordinates, apparent magnitude, distance, proper motion (p.m.), spectral-type, and physical properties of each host star. Near-infrared magnitudes for the primary star are from 2MASS \citep{skrutskie_06}. Distance estimates are based on measured parallax from the {\it Hipparcos} satellite \citep{van_leeuwen_07}. Effective temperature ($T_{\rm eff}$), surface gravity (log g), and metallicity ($[\mbox{Fe/H}]$) are derived using SME. Ages are estimated using theoretical isochrones for HD~53665 ($B-V=0.47$) \citep{valenti_fischer_05}, and gyrochronology for HD~68017 ($B-V=0.69$) and HD~71881 ($B-V=0.63$) \citep{mamajek_hillenbrand_08}. (Bottom) Companion magnitude difference, absolute magnitude, estimated spectral-type, mass constraint from dynamics ($m_{dyn}$) and estimated mass from photometry using the \citealt{dotter_08} atmospheric models ($m_{\rm model}$) and \citealt{delfosse_00} $M_K$--mass empirical relations ($m_{\rm empirical}$).}
\label{tab:starprops}
\end{table*}

\section{COMPANION MASS ESTIMATE}\label{sec:mass}
\subsection{Mass from Photometry}\label{sec:phot}
We measured the brightness of each companion relative to its host star by performing aperture photometry, accounting for contamination from the primary. Stellar magnitudes were first converted from 2MASS measurements \citep{2MASS_06} to the MKO filter system \citep{tokunaga_02} using transformations from \citealt{carpenter_01}.\footnote{http://www.astro.caltech.edu/~jmc/2mass/v3/transformations/} Our observations were acquired in the $J, H, K'$ (MKO) bands. We have assumed that $\Delta K \approx \Delta K'$, an assumption that is justified given the relatively mild colors of solar-type stars, and fact that uncertainty in the measured magnitude difference and parallax dominate the uncertainty in absolute magnitude. The mass of each companion is found by comparing its brightness to late-type dwarfs using \citealt{dotter_08} theoretical evolutionary tracks (the Dartmouth models), and also the \citealt{delfosse_00} empirical relations which correlate $M_K$ with mass. Differential, apparent, and absolute magnitudes are listed in Table 2. 

We find that HD~53665~B has a mass of $0.65\pm0.03M_{\odot}$ based on photometry \citep{dotter_08}. Its absolute magnitude in each near-infrared band is consistent with this value. Using Table~5 from \citealt{kraus_hillenbrand_07}, HD~53665~B has colors and brightness consistent with either an K7 or M0 dwarf. For comparison, we find a mass of $0.70\pm0.03M_{\odot}$ based on the measured $M_{K'}=4.81\pm0.16$ \citep{delfosse_00}.

HD~68017~B has a mass of $0.16\pm0.02M_{\odot}$ based on photometry \citep{dotter_08}. Its absolute magnitude is estimated only in the K-band, as our unocculted H-band observations of the primary star from Jan. 7, 2012 were saturated. Table~2 shows a lower-limit for the H-band magnitude. HD~68017~B has a K-band brightness consistent with an M5-dwarf \citep{kraus_hillenbrand_07}. For comparison, we find a mass of $0.15\pm0.01M_{\odot}$ based on the measured $M_{K'}=8.31\pm0.11$ \citep{delfosse_00}.

HD~71881~B has a mass of $0.31\pm0.03M_{\odot}$ based on photometry. Its absolute magnitude in the H and K bands are both consistent with this value. HD~71881~B has colors and brightness most consistent with either an M3 or M4 dwarf \citep{kraus_hillenbrand_07}. For comparison, we find a mass of $0.29\pm0.02M_{\odot}$ based on the measured $M_{K'}=6.99\pm0.11$ \citep{delfosse_00}.
%The estimated mass however corresponds more closely to an M3-dwarf. We adopt an intermediate designation of M3.5V.

\subsection{Mass Lower-limit from Dynamics}\label{sec:dyn}
When combined with a RV trend, a single epoch of imaging observations (single measurement of the projected physical separation) places a lower-limit on the companion dynamical mass \citep{torres_99,liu_02}. We have measured the instantaneous Doppler acceleration for HD~53665 and HD~71881, assuming the RV data may be approximated as linear across the full time baseline. Using a Markov chain Monte Carlo (MCMC) analysis, we find slopes of:
%the HD~68017 RV curve was measured using the most recent 21 data points, from Sept. 25, 2010 to present. Using an MCMC analysis, we find slopes: 
\begin{eqnarray}
\left(dv/dt \right)_{\mbox{\tiny{HD53665}}} &=& \; \: +5.3\pm0.3   \; \mbox{m} \: \mbox{s}^{-1} \mbox{yr}^{-1}  \nonumber \\
%\left(dv/dt \right)_{\mbox{\tiny{HD68017}}} &=&   +16.3\pm0.9  \;  \mbox{m} \: \mbox{s}^{-1} \mbox{yr}^{-1}  \nonumber \\
\left(dv/dt \right)_{\mbox{\tiny{HD71881}}} &=& -10.3\pm0.2 \; \mbox{m} \: \mbox{s}^{-1} \mbox{yr}^{-1}.        \nonumber
\end{eqnarray} 
With a projected separation of 102.9 AU (Feb. 22, 2011), HD~53665~B has a minimum dynamical mass of $0.77\pm0.14M_{\odot}$. Accounting for uncertainty in the stellar parallax, measured angular separation, and RV acceleration, we adopt a minimum mass of $0.63M_{\odot}$. This value is consistent with the mass derived from photometry provided HD~53665~B has an edge-on orbit (modulo faulty assumptions regarding system coevolution, or systematic errors in the isochrone models). Likewise,  HD~71881~B has a projected separation of 35.2 AU (Feb. 22, 2011), which corresponds to a minimum dynamical mass of $0.17M_{\odot}$. Constraints from Doppler RV and imaging data are compared to mass estimates from photometry in Table~2.

HD~68017~B has a projected separation of $13.0^{+0.1}_{-0.2}$ AU (Feb. 22, 2011). Evaluating the local RV slope using the most recently obtained 21 data points (from Sept. 25, 2010 to present), we find $\left(dv/dt \right)_{\mbox{\tiny{HD68017}}} =+16.3\pm0.9  \;  \mbox{m} \: \mbox{s}^{-1} \mbox{yr}^{-1}$, which corresponds to a minimum dynamical mass of $0.03M_{\odot}$. However, the full RV time series shows significant curvature (change in the acceleration), enabling a more sophisticated analysis. We have performed MCMC simulations that simultaneously fit the Doppler and astrometric data using a Keplerian orbit. Our calculations account for the measured orbital motion of the companion from both imaging epochs. While the currently available data set provides insufficient information to calculate a unique orbit inclination, we find that the companion minimum mass constraint becomes $m_{\rm dyn}>0.08M_{\odot}$ ($68.2\%$ confidence). 

\section{SUMMARY}
We have established a new observing program that uses precise RV measurements to identify promising targets for high-contrast imaging observations. The nearby, solar-type stars HD~53665, HD~68017, and HD~71881 exhibit long-term Doppler accelerations (``trends"). We have used NIRC2 AO observations at Keck to directly detect the companions responsible for causing the trend in each case. Follow-up astrometry demonstrates that each candidate is co-moving with the primary star. Relative photometry measurements suggest spectral-types of K7--M0, M5, and M3--M4 respectively. 

% \footnote{Our Doppler measurements provide more than a decade of time baseline leading up to the direct imaging discovery.}
As inferred from their host star, each M-dwarf companion is a metallicity and age benchmark object. With continued Doppler monitoring and follow-up AO observations, two out of the three companions, HD~68017~B and possibly HD~71881~B, will also serve as mass benchmarks.  HD~68017 has a projected separation of only $13.0^{+0.1}_{-0.2}$ AU and already shows significant RV curvature (change in the acceleration) and measurable astrometric motion, making it possible to estimate a dynamical mass in the next several years.\footnote{For comparison, given the typical (high) signal-to-noise ratio of Doppler measurements for solar-type stars, and direct astrometric measurements, \citealt{crepp_12a} have shown it possible to calculate accurate dynamical masses for companions with semimajor axes as large as $\approx$19 AU.} Using currently available data, we calculate a lower-limit to the mass of each companion. We posit that HD~53665~B must have a near edge-on orbit given the agreement between the mass estimate from photometry and dynamics. 

Few mass, age, and metallicity benchmark dwarfs are known to date \citep{liu_07,dupuy_10,bowler_12,bowler_12b}. The goal of the TRENDS high-contrast imaging program is to discover and characterize low-mass stellar and substellar companions with physical properties determined independently from spectro-photometric measurements, in order to calibrate theoretical atmospheric models and thermal evolutionary models. Each of the companions presented are amenable to direct spectroscopy using AO-fed integral-field units, such as OSIRIS at Keck or Project 1640 at Palomar. 

\section{ACKNOWLEDGEMENTS}
JC acknowledges support from NASA Origins grant NNX13AB03G. The data presented herein were obtained at the W.M. Keck Observatory, which is operated as a scientific partnership among the California Institute of Technology, the University of California and the National Aeronautics and Space Administration. The Observatory was made possible by the generous financial support of the W.M. Keck Foundation.

\begin{small}
\bibliographystyle{jtb}
\bibliography{ms.bib}
\end{small}

\begin{deluxetable}{lll}
\tablecaption{Radial Velocities for HD 53665\label{vel53665}}
\tablewidth{0pt}
\tablehead{
\colhead{HJD} &
\colhead{RV} &
\colhead{Uncertainty} \\
\colhead{-2,440,000} &
\colhead{(m~s$^{-1}$)} &
\colhead{(m~s$^{-1}$)} 
}
\startdata
10838.8830 &  -25.83 &  5.69 \\
10861.8568 &   -9.69 &  5.98 \\
11071.1199 &  -16.24 &  5.84 \\
11171.9255 &  -14.63 &  5.79 \\
11226.8781 &  -17.12 &  5.65 \\
11544.0550 &   -1.81 &  5.92 \\
11552.9643 &  -20.22 &  6.06 \\
11585.9500 &  -23.16 &  5.82 \\
11883.0298 &  -33.55 &  5.82 \\
11973.8361 &  -22.17 &  5.91 \\
12235.9597 &  -16.01 &  5.60 \\
12574.0727 &    0.74 &  5.70 \\
12600.9894 &   -7.03 &  5.60 \\
12653.0320 &   15.80 &  5.83 \\
12987.9295 &  -13.35 &  5.77 \\
13304.1114 &   16.06 &  5.24 \\
13400.9519 &   22.45 &  5.38 \\
14024.1377 &   31.39 &  5.45 \\
15521.9975 &   32.68 &  5.38 \\
15672.8256 &   57.91 &  5.50 \\
15871.0233 &   43.77 &  5.25
\\
\enddata
\end{deluxetable}

\begin{deluxetable}{lll}
\tablecaption{Radial Velocities for HD 68017\label{vel68017}}
\tablewidth{0pt}
\tablehead{
\colhead{HJD} &
\colhead{RV} &
\colhead{Uncertainty} \\
\colhead{-2,440,000} &
\colhead{(m~s$^{-1}$)} &
\colhead{(m~s$^{-1}$)} 
}
\startdata
10461.9269 &   50.69 &  2.36 \\
10784.1169 &   41.84 &  2.41 \\
10807.1673 &   30.26 &  2.74 \\
10837.8347 &   39.68 &  2.46 \\
10838.9684 &   35.12 &  2.53 \\
10861.8262 &   39.34 &  2.46 \\
10862.7586 &   42.87 &  2.38 \\
11171.0900 &   18.20 &  2.44 \\
11226.7940 &   31.60 &  2.51 \\
11227.9052 &   30.02 &  2.43 \\
11229.9098 &   32.27 &  2.33 \\
11551.0555 &   17.13 &  2.61 \\
11583.9188 &   15.21 &  2.46 \\
11884.1426 &   14.94 &  2.52 \\
11898.1321 &   21.48 &  2.50 \\
11899.1205 &   23.57 &  2.41 \\
11900.0979 &   18.66 &  2.33 \\
11901.1537 &   16.07 &  2.40 \\
11972.0135 &   11.99 &  2.44 \\
11972.9874 &   15.01 &  2.39 \\
11973.8854 &   16.30 &  2.45 \\
11974.8692 &   12.93 &  2.49 \\
12243.0488 &    5.48 &  2.58 \\
12573.1006 &    2.04 &  2.53 \\
12601.0207 &   -2.07 &  2.44 \\
12680.9684 &   -8.36 &  2.47 \\
12711.7407 &   -2.32 &  2.58 \\
12988.0081 &  -12.00 &  2.55 \\
13303.1395 &   -6.03 &  2.78 \\
13398.8550 &  -15.14 &  2.12 \\
13692.9970 &  -14.80 &  2.13 \\
13693.0957 &  -16.00 &  2.12 \\
13694.1285 &  -17.28 &  2.13 \\
13695.1055 &  -16.35 &  2.13 \\
13696.0847 &  -16.56 &  2.19 \\
13697.1010 &  -28.71 &  2.18 \\
13724.0212 &  -17.79 &  2.15 \\
13725.0429 &  -18.73 &  2.15 \\
13747.0478 &  -12.93 &  2.14 \\
13747.9616 &  -15.76 &  2.15 \\
13748.9262 &  -16.15 &  2.37 \\
13749.8534 &  -13.49 &  2.16 \\
13750.8602 &  -15.83 &  2.15 \\
13751.9266 &  -10.32 &  2.39 \\
13752.9635 &  -14.96 &  2.14 \\
13753.9402 &  -12.69 &  2.13 \\
13775.7943 &  -13.93 &  2.14 \\
13776.9239 &  -10.81 &  2.14 \\
14129.9829 &  -16.35 &  2.14 \\
14399.1317 &  -10.94 &  3.11 \\
14492.9688 &  -22.14 &  2.43 \\
14547.9008 &  -16.94 &  2.46 \\
14548.8355 &  -11.53 &  2.53 \\
14809.9817 &   -6.87 &  2.59 \\
14867.8982 &   -8.70 &  2.48 \\
14927.8853 &  -12.18 &  2.56 \\
14929.8538 &   -2.49 &  2.48 \\
14984.7583 &   -8.21 &  2.42 \\
15109.1438 &   -5.35 &  3.02 \\
15134.1513 &    0.84 &  3.02 \\
15164.0166 &    1.78 &  2.44 \\
15172.1149 &   -4.52 &  2.52 \\
15188.0140 &   -2.65 &  2.51 \\
15189.9820 &   -1.62 &  2.44 \\
15192.0018 &   -0.65 &  2.49 \\
15198.9722 &   -8.83 &  2.50 \\
15229.1011 &   -1.10 &  2.16 \\
15229.7942 &   -5.34 &  2.17 \\
15231.8098 &    2.19 &  2.17 \\
15251.9199 &    0.16 &  2.17 \\
15255.7629 &   -5.62 &  2.15 \\
15260.7790 &   -6.72 &  2.16 \\
15289.7242 &    2.54 &  2.50 \\
15311.7447 &    3.50 &  2.49 \\
15344.7419 &   -6.89 &  2.69 \\
15465.1494 &    5.98 &  2.36 \\
15468.1055 &    4.81 &  2.36 \\
15469.1528 &    3.65 &  2.37 \\
15470.1502 &    9.43 &  2.36 \\
15487.1576 &    5.54 &  2.35 \\
15490.1555 &    8.16 &  2.41 \\
15491.1524 &    8.56 &  2.35 \\
15501.1589 &    9.99 &  2.38 \\
15522.0236 &    2.12 &  2.27 \\
15543.1041 &    3.73 &  2.16 \\
15605.9635 &   16.44 &  2.19 \\
15634.8186 &    8.89 &  2.16 \\
15671.7972 &   14.31 &  2.13 \\
15697.7307 &   12.49 &  2.17 \\
15698.7813 &   11.57 &  2.18 \\
15843.0775 &   24.08 &  2.38 \\
15880.1365 &   28.22 &  2.93 \\
15904.1782 &   22.53 &  2.58 \\
15960.9100 &   33.29 &  2.55 \\
15999.7837 &   27.41 &  2.52 \\
16073.7601 &   29.82 &  2.49
\\
\enddata
\end{deluxetable}

\begin{deluxetable}{lll}
\tablecaption{Radial Velocities for HD 71881\label{vel71881}}
\tablewidth{0pt}
\tablehead{
\colhead{HJD} &
\colhead{RV} &
\colhead{Uncertainty} \\
\colhead{-2,440,000} &
\colhead{(m~s$^{-1}$)} &
\colhead{(m~s$^{-1}$)} 
}
\startdata
10807.1721 &  112.06 &  2.56 \\
10837.9895 &  106.74 &  2.37 \\
10862.7614 &  109.55 &  2.46 \\
10862.8804 &  108.43 &  2.37 \\
11171.0933 &   97.73 &  2.35 \\
11227.9372 &   98.27 &  2.32 \\
11552.0265 &   79.55 &  2.42 \\
11900.1047 &   76.05 &  2.33 \\
12003.9078 &   69.54 &  2.47 \\
12007.8729 &   66.09 &  2.39 \\
12062.7530 &   69.57 &  2.47 \\
12064.7717 &   71.06 &  2.55 \\
12236.0510 &   56.62 &  2.54 \\
12573.1159 &   51.72 &  2.40 \\
12712.8234 &   48.80 &  2.50 \\
12988.0181 &   45.16 &  2.43 \\
13303.1333 &   50.16 &  3.91 \\
13425.8843 &   25.04 &  4.67 \\
15556.1133 &   -8.53 &  4.41 \\
15871.0606 &  -34.85 &  4.41 \\
16018.8872 &  -40.97 &  4.52
\\
\enddata
\end{deluxetable}

\end{document}